\documentclass[10pt,journal]{IEEEtran}

\usepackage{graphicx} 
\usepackage{float}
\usepackage{cite}
\usepackage{amsmath,amssymb}
\usepackage[outdir=./]{epstopdf}
\usepackage{afterpage}
\usepackage{optidef}
\usepackage{orcidlink}
\usepackage{soul}
\usepackage{tikz}
\usepackage{pgfplots}
\usetikzlibrary{shapes.geometric, fit, positioning, backgrounds}
\usepackage{multirow}
\usepackage{subcaption}

\usepackage[firstpage=true]{background}
\SetBgContents{\textcolor{black}{\footnotesize This work has been submitted to the IEEE for possible publication. Copyright may be transferred without notice, after which this version may no longer be accessible.}}
\SetBgPosition{current page.south}
\SetBgVshift{0.50cm}
\SetBgOpacity{1.0}
\SetBgAngle{0.0}
\SetBgScale{1.0}

\makeatletter
\ams@newcommand{\iiiiint}{\DOTSI\protect\MultiIntegral{5}}
\renewcommand{\MultiIntegral}[1]{%
  \edef\ints@c{\noexpand\intop
    \ifnum#1=\z@\noexpand\intdots@\else\noexpand\intkern@\fi
    \ifnum#1>\tw@\noexpand\intop\noexpand\intkern@\fi
    \ifnum#1>\thr@@\noexpand\intop\noexpand\intkern@\fi
    \ifnum#1>4 \noexpand\intop\noexpand\intkern@\fi 
    \noexpand\intop
    \noexpand\ilimits@
  }%
  \futurelet\@let@token\ints@a
}
\makeatother

\begin{document}
\bstctlcite{IEEEexample:BSTcontrol}

\title{Semantic Communication for Cooperative Multi-Task Processing over Wireless Networks}

\author{
\IEEEauthorblockN{Ahmad Halimi Razlighi\,\orcidlink{0009-0006-3826-832X}, Carsten Bockelmann\,\orcidlink{0000-0002-8501-7324}, and Armin Dekorsy\,\orcidlink{0000-0002-5790-1470}} \\

\IEEEauthorblockA{Department of Communications Engineering, University of Bremen, Germany}\\

\IEEEauthorblockA{E-mails:\{halimi, bockelmann, dekorsy\}@ant.uni-bremen.de}
}

\maketitle

\begin{abstract}

In this paper, we investigated semantic communication for multi-task processing using an information-theoretic approach. We introduced the concept of a ``semantic source", allowing multiple semantic interpretations from a single observation. We formulated an end-to-end optimization problem taking into account the communication channel, maximizing mutual information (infomax) to design the semantic encoding and decoding process exploiting the statistical relations between semantic variables. To solve the problem we perform data-driven deep learning employing variational approximation techniques. Our semantic encoder is divided into a common unit and multiple specific units to facilitate cooperative multi-task processing. Simulation results demonstrate the effectiveness of our proposed semantic source and system design when statistical relationships exist, comparing cooperative task processing with independent task processing. However, our findings highlight that cooperative multi-tasking is not always beneficial, emphasizing the importance of statistical relationships between tasks and indicating the need for further investigation into the semantically processing of multiple tasks.

\end{abstract}


\section{Introduction} \label{section.Intro}
Applications involving machine-to-machine or human-to-machine communications often have to prioritize task execution over the exact reconstruction of transmitted information at the receiver. Unlike the traditional information theory established by Shannon, which emphasizes the accurate transmission and reception of bits, the design of communication systems for these applications takes a distinct approach, drawing attention to task performance rather than fidelity in information transmission. Thus, regarding the three levels of communication \cite{Sana2022}, technical layer (accurate transmission of symbols), semantic level (transmitting the desired meaning), and effectiveness level (effectiveness of the received meaning), one should investigate the second level of communication to meet the demands of emerging applications. Leveraging advancements in artificial intelligence, deep learning, and end-to-end (E2E) communication technologies, the concept of \emph{semantic communication} has recently emerged \cite{Gunduz2022}. Semantic communication prioritizes understanding the meaning and goals behind transmitted information, surpassing the traditional focus on the precise transmission of bits.

Four approaches to semantic communication are outlined in \cite{Wheeler2023}. Firstly, the classical approach utilizes \emph{logical} probability to quantify semantic information, primarily for text sources. Secondly, the knowledge graph (KG) approach represents semantics by KG structure. Thirdly, the machine learning (ML) approach leverages learned model parameters to represent semantics. Lastly, the significance approach emphasizes \emph{timing} as semantics. Inspired by Weaver, an alternative approach extends Shannon's \emph{statistical} probability (information theory) beyond the technical layer to the next two levels \cite{Weaver}.

Works in semantic communication have been categorized into two types of research directions: data reconstruction and task execution. Data reconstruction is generally done by extracting semantic information on the transmitter side and recovering data using the received semantic information on the receiver side \cite{Xie2021}, \cite{Xie2021-2}, \cite{Yan2022}, \cite{Tong2021}.

On the other hand, for task execution, also known as task-oriented communication, \cite{Shao2021} explored a communication scheme based on the information bottleneck (IB) framework, enabling information encoding for a single task while adapting to dynamic wireless channel conditions. The same authors in \cite{Shao2022} studied distributed relevant information encoding for collaborative feature extraction to fulfill a single task. Moreover, \cite{Beck2023} offered a framework for the collaborative recovery of the semantics as a single task using encoded information of distributed sources and expanded it using reinforcement learning in \cite{beck2024modelfree}. \cite{Binucci} has contributed to resource allocation in a multi-user system according to the single task accuracy, channel conditions, and computing requests.

On multi-task processing in semantic communication, \cite{xie2022task} and \cite{he2022learning} explored non-cooperative methods where each task operates on its respective dataset independently. Conversely, recent works like \cite{10013075}, \cite{10520522}, and \cite{gong2023scalable} studied joint multi-tasking using established ML approaches and architecture \cite{caruana1997multitask}, for semantic communication. Although these works incorporated communication aspects like channel conditions in their studies, their multi-task processing is based exclusively on ML approaches.

Existing research on multi-tasking in semantic communication lacks information-theoretic analysis of the problem, and task-oriented communication works have only addressed single-task processing based on information theory. We aim to advance task-oriented research direction in multi-task processing for semantic communication based on information theory and provide new insights. Our information-theoretic evaluations employ a data-driven approach and variational approximations. In addition, our evaluations reveal the alignment of our split structure with the neural network (NN) structures in multi-tasking of existing ML works, such as \cite{caruana1997multitask}, by indicating that the first encoder should be shared amongst the specific ones. Compared to existing literature, our approach is distinct since we have tailored multi-task processing to semantic communication through information theory avoiding black-box use of NNs. Key contributions include:

\begin{itemize}
    \item Modeling semantic source utilizing probabilistic modeling to enable multiple different semantics extraction of a single observation at the same time.
    \item Studying a semantic encoding and decoding structure employing NNs and considering the wireless channel based on an information-theoretic analysis, wherein the encoder is divided into a common unit (CU) and multiple specific units (SUs).
    \item Demonstrating that the structure enhances task execution performance through cooperation in the CU when a statistical relationship exists between tasks, compared to when SU encoders and decoders operate independently without the CU.
    \item Highlighting that cooperative multi-task processing is not always beneficial, justified by the statistical relationship between semantic variables, which opens new research questions for further investigation.
\end{itemize}

\section{System Model} \label{section.SystemModel}

In this section, we introduce our probabilistic modeling of a semantic source and corresponding system model. Next, we formulate an information-theoretical optimization problem to address the execution of multiple tasks.

\begin{figure}[!t]
    \centering
    \scalebox{0.9}{
    \begin{tikzpicture}[>=stealth, node distance=1.5cm]
        \node[text=red] (G) {};
        \node[right=3 of G, text=red] (Z) {};
        \node[right=0.32 of G] (B) {};
        \node[left=0.3 of Z] (Q) {};
        \node[right=2.5 of Z, yshift=0.3 cm] (S) {$\begin{tabular}{@{}c@{}}
                                                    \text{Observation} \\
                                                    \small \text{($\mathbf{S}$)}
                                                  \end{tabular}$};
        \draw[fill=white] (G.center)++(0.8,0.8) circle (0.5) node[font=\scriptsize] (GCn) {$\text{Task}_N$};
        \draw[fill=white] (Z.center)++(0.8,0.8) circle (0.5) node[font=\small] (ZCn) {$z_N$};
        \node[right=0.13 of G, xshift=0.8 cm, yshift=0.8 cm] (Bn) {};
        \node[left=0.1 of Z, xshift=0.8 cm, yshift=0.8 cm] (Qn) {};
        \draw[->] (Bn) -- (Qn);
        \draw[fill=white] (G.center)++(0.3,0.3) circle (0.6) node (GC3) {};
        \draw[fill=white] (Z.center)++(0.3,0.3) circle (0.6) node (ZC3) {};
        \node[right=0.23 of G, xshift=0.3 cm, yshift=0.3 cm] (B2) {};
        \node[left=0.3 of Z, xshift=0.3 cm, yshift=0.3 cm] (Q2) {};
        \draw[->] (B2) -- (Q2);
        \draw[fill=white] (G.center)++(0.15,0.15) circle (0.65) node (GC2) {};
        \draw[fill=white] (Z.center)++(0.15,0.15) circle (0.65) node (ZC2) {};
        \node[right=0.28 of G, xshift=0.15 cm, yshift=0.15 cm] (B1) {};
        \node[left=0.3 of Z, xshift=0.15 cm, yshift=0.15 cm] (Q1) {};
        \draw[->] (B1) -- (Q1); 
        \draw[fill=white] (G.center) circle (0.7) node[font=\small] (GC1) {$\text{Task 1}$};
        \draw[fill=white] (Z.center) circle (0.7) node[font=\large] (ZC1) {$z_1$};
        \draw[->] (B) -- (Q);

        \fill (G.center)++(2.0,0.5) circle [radius=0.7pt];
        \fill (G.center)++(2.07,0.57) circle [radius=0.7pt];
        \fill (G.center)++(2.14,0.64) circle [radius=0.7pt];

        \fill (G.center)++(5.05,0.5) circle [radius=0.7pt];
        \fill (G.center)++(5.12,0.57) circle [radius=0.7pt];
        \fill (G.center)++(5.19,0.64) circle [radius=0.7pt];

        \node[right=0.32 of Z] (QQ) {};
        \node[right=2.47 cm of Z] (BB) {};
        \draw[->] (QQ) -- (BB);
        \node[right=0.29 of Z, xshift=0.15 cm, yshift=0.15 cm] (QQ1) {};
        \node[right=2.33 of Z, xshift=0.15 cm, yshift=0.15 cm] (BB1) {};
        \draw[->] (QQ1) -- (BB1);
        \node[right=0.24 of Z, xshift=0.3 cm, yshift=0.3 cm] (QQ2) {};
        \node[right=2.18 of Z, xshift=0.3 cm, yshift=0.3 cm] (BB2) {};
        \draw[->] (QQ2) -- (BB2);
        \node[right=0.13 of Z, xshift=0.8 cm, yshift=0.8 cm] (QQn) {};
        \node[right=1.68 of Z, xshift=0.8 cm, yshift=0.8 cm] (BBn) {};
        \draw[->] (QQn) -- (BBn);
        \node[draw, rectangle, fit=(S), inner sep=1pt, minimum height=2 cm] (RectangleS) {};
        \node[left=0.35 of Z] (ZSS) {};
        \node[draw, dashed, fit=(RectangleS)(ZSS), inner sep=5pt] (SS) {};
        \node[above=0.1cm of SS, font=\small] {Semantic Source};
    \end{tikzpicture}
    }
    \caption{Probabilistic graphical modeling of the semantic source.}
    \label{fig:semantic_source_model}
\end{figure}

\subsection{Semantic Source Modeling}\label{subsec:semantic_source}
We model our \emph{semantic source},  as shown in Fig. \ref{fig:semantic_source_model}. We assume the existence of $N$ independent tasks. Each task is entailed with its specific \emph{semantic variable}, thus we have $N$ independent semantic variables indicated by $\mathbf{z}=[\,z_1\, z_2\,\dots\,z_N]\,$. We assume that our semantic variables are entailed with observation, $\mathbf{S}$, and consider these entailments stochastic \cite{Bao}. We define the tuple of $(\mathbf{z}, \mathbf{S})$ as our semantic source, fully described by the probability distribution of $p(\mathbf{z}, \mathbf{S})$. More precisely, we describe our semantic source by $p(\mathbf{z})p(\mathbf{S}|\mathbf{z})$, where $p(\mathbf{S}|\mathbf{z})$ is our \emph{semantic channel} that reflects the semantic variables in our observation. Such a definition enables the simultaneous extraction of multiple semantic variables based on a single observation and addresses multiple tasks. For instance, consider an image featuring both a tree and a number. One task may entail determining the presence of a tree, resulting in a binary semantic variable. Meanwhile, another task could focus on identifying the number within the image, yielding a multinomial semantic variable, independent of the first one.

\subsection{System Probabilistic Modeling}\label{subsec.system_prob_model}
When some statistical relationship exists between semantic variables, there can be some common relevant information useful for them. Thus, our idea is to split up the semantic encoder into a CU and multiple SUs, introducing cooperative semantic communication to leverage the common information and process multiple tasks. Thus, we aim to offer an information-theoretic design and investigation of a multi-tasking structure for semantic communication. Our system model consists of a single observation and $N$ independent semantic variables, each associated with a unique task. As illustrated in Fig. \ref{fig:system-model} initially, the CU encoder extracts the common relevant information from the semantic source. Then, $N$ SU encoders extract and transmit task-specific information to their respective decoders. Output of SU encoders are shown as $\mathbf{x}_1, \mathbf{x}_2,\dots,\mathbf{x}_N$, and their noise-corrupted version received at the corresponding decoders are indicated by $\mathbf{\hat{x}}_1, \mathbf{\hat{x}}_2,\dots,\mathbf{\hat{x}}_N$. 

Our approach uses the additive white Gaussian noise (AWGN) channel to incorporate wireless transmission between encoders and decoders. Upon reception, semantic decoders deliver the semantic variables to their respective recipients.
Specifically, the Markov representation of our system model for the i-th semantic variable is outlined as follows.

\begin{equation}
\begin{split}
    &p(\hat{z}_i,\mathbf{\hat{x}}_i,\mathbf{x}_i,\mathbf{c}|\mathbf{S}) =\\[0.3em] & p^{\text{\tiny Dec$_i$}}(\hat{z}_i|\mathbf{\hat{x}}_i)\,p^{\text{\tiny Channel}}(\mathbf{\hat{x}}_i|\mathbf{x}_i)\,p^{\text{\tiny SU$_i$}}(\mathbf{x}_i|\mathbf{c})p^{\text{\tiny CU}}(\mathbf{c}|\mathbf{S}).
\end{split}
\label{eq:system_probability}
\end{equation}

In (\ref{eq:system_probability}), $p^{\text{\tiny CU}}(\mathbf{c}|\mathbf{S})$ defines the CU that extracts the common relevant information amongst all tasks, from the observation. The i-th SU is described by $p^{\text{\tiny SU$_i$}}(\mathbf{x}_i|\mathbf{c})$ extracting task-specific information and provide $\mathbf{x}$ as the channel input. The corresponding decoder is then specified by $p^{\text{\tiny Dec$_i$}}(\hat{z}_i|\mathbf{\hat{x}}_i)$, where $\mathbf{\hat{x}}$ is the received information passed through the AWGN channel and modeled like $\mathbf{\hat{x}}_i = \mathbf{x}_i + \mathbf{n}$, where $\mathbf{n} \sim \mathcal{N}(\mathbf{0}_m, \sigma^2_n \mathbf{I}_m)$, and $m$ is the size of the encoded task-specific information.
\begin{figure}[!t]
    \centering
    \scalebox{0.9}{
    \begin{tikzpicture}
        \node[draw, circle, inner sep=0.5pt] (S) {$\begin{tabular}{@{}c@{}}
                                                    \text{Semantic} \\
                                                    \text{Source} \\
                                                    \text{$(\mathbf{z}, \mathbf{S})$}
                                                  \end{tabular}$};
        \node[draw, rectangle, right=0.5 cm of S, inner sep=4pt] (C) {$\text{CU Enc.}$};
        \node[below=0.3cm of C] (th) {$\boldsymbol{\theta}$};
        \node[draw, rectangle, right=0.3 cm of C, yshift=1.2 cm, inner sep=4pt] (SU1) {$\text{SU Enc.}$};
        \node[below=0.3cm of SU1] (ph1) {$\boldsymbol{\phi}_1$};
        \node[draw, rectangle, dashed, fit=(SU1)(ph1), inner sep=1.2pt](outerSU1){};
        \node[draw, rectangle, right=0.3 cm of C, yshift=-1.2 cm, inner sep=4pt] (SU2) {$\text{SU Enc.}$};
        \node[below=0.3cm of SU2] (ph2) {$\boldsymbol{\phi}_N$};
        \node[draw, dotted, line width=1pt, fit=(SU2)(ph2), inner sep=1.2pt](outerSU2){};
        \node[right=0.4 cm of SU1](X1){$\mathbf{x}_1$};
        \node[right=0.4 cm of SU2](Xn){$\mathbf{x}_N$};
        \node[right=of SU2, yshift=1 cm, xshift=-0.02 cm] (ch) {$\begin{tabular}{@{}c@{}}
              \footnotesize C\\
              \footnotesize h\\
              \footnotesize a\\
              \footnotesize n\\
              \footnotesize n\\
              \footnotesize e\\
              \footnotesize l
        \end{tabular}$};
        \node[draw, fit=(ch), minimum height=3.5 cm, inner sep=-2pt]{};
        
        \node[draw, rectangle, right=2.5 of SU1] (Rx1) {$\text{Dec.}$};
        \node[below=0.3cm of Rx1] (ps1) {$\boldsymbol{\psi}_1$};
        \node[draw, dashed, fit=(Rx1)(ps1), inner sep=1.3pt] (RxSU1) {};
        \node[draw, rectangle, right=2.5 of SU2] (Rx2) {$\text{Dec.}$};
        \node[below=0.3cm of Rx2] (ps2) {$\boldsymbol{\psi}_N$};
        \node[draw, dotted, line width=1pt, fit=(Rx2)(ps2), inner sep=1.3pt] (RxSU2) {};
        \node[left=0.4 cm of Rx1](X1_hat){$\mathbf{\hat{x}}_1$};
        \node[left=0.4 cm of Rx2](Xn_hat){$\mathbf{\hat{x}}_N$};
        \node[right=0.4 cm of Rx1](Z1_hat){$\hat{z}_1$};
        \node[right=0.4 cm of Rx2](Zn_hat){$\hat{z}_N$};

        \node[below=0.5 cm of S] (task1) {$\text{Task 1}$};
        \node[below=0.05 cm of task1] (task2) {$\text{Task N}$};
        \node[right=of task1, inner sep=0pt] (ta1) {};
        \node[right=of task2, inner sep=0pt] (ta2) {};
        \node[above=0.03 cm of C, xshift= 0.7 cm]{$\mathbf{c}$};

        \draw[dotted, line width=2pt] (outerSU1) -- (outerSU2);
        \draw[dotted, line width=2pt] (RxSU1) -- (RxSU2);
        \draw[->, line width=1pt] (S) -- node[above]{\small$\mathbf{S}$} (C.west);
        \draw[->, line width=1pt] (C.east) -- (SU1.west);
        \draw[->, line width=1pt] (C.east) -- (SU2.west);
        \draw[->, line width=1pt] (SU1) -- (X1);
        \draw[->, line width=1pt] (SU2) -- (Xn);
        \draw[->, line width=1pt] (X1_hat) -- (Rx1);
        \draw[->, line width=1pt] (Xn_hat) -- (Rx2);
        \draw[->, line width=1pt] (Rx1) -- (Z1_hat);
        \draw[->, line width=1pt] (Rx2) -- (Zn_hat);
        \draw[->, line width=1pt] (th) -- (C);
        \draw[->, line width=1pt] (ph1) -- (SU1);
        \draw[->, line width=1pt] (ph2) -- (SU2);
        \draw[->, line width=1pt] (ps1) -- (Rx1);
        \draw[->, line width=1pt] (ps2) -- (Rx2);
        \draw[draw, dashed] (task1) -- (ta1);
        \draw[draw, dotted, line width=1pt] (task2) -- (ta2);
        
    \end{tikzpicture}
    }
    \caption{Illustration of the cooperative multi-task semantic communication system model.}
    \label{fig:system-model}
\end{figure}

\subsection{Optimization Problem} \label{subsec.Loss_function}

To design our split semantic encoder architecture, we formulate an optimization problem adopting the information maximization principle together with the E2E learning manner, which has been proven effective for task-oriented communication \cite{9145068}, as follows.

\begin{equation} \label{eq:optimization_problem}
        \left[\,p^{\text{\tiny CU}}(\mathbf{c}|\mathbf{S})^\star,\, p^{\text{\tiny SU}}(\mathbf{x}|\mathbf{c})^\star\right]\ = \arg \underset{\substack{p^{\text{\tiny CU}}(\mathbf{c}|\mathbf{S}), \\ p^{\text{\tiny SU}}(\mathbf{x}|\mathbf{c}).}}{\text{max}}\, \sum_{i=1}^{N}\, b_i\, I(\mathbf{\hat{x}}_i;z_i).
\end{equation}

Thus, the objective is to maximize the mutual information between the channel output $\mathbf{\hat{x}}_i$, and the semantic variables $z_i$. In equation (\ref{eq:optimization_problem}), $b_i$ is a constant coefficient, representing a factor that will be fixed at one. This choice is made as we do not explore the relationship between semantic variables or prioritize them within the scope of this paper.
Expanding the mutual information in our optimization problem as discussed in detail in Appendix \ref{Appendix:derivation}, the approximated objective function is derived like:
\begin{equation} \label{eq:objective_function}
    \begin{aligned}
        &\mathcal{L}(\boldsymbol{\theta}, \boldsymbol{\Phi}) = \sum_{i=1}^{N}\, I(\mathbf{\hat{x}}_i; z_i) \\[0.6em]
        &\approx \underbrace{ \textcolor{black}{\mathbb{E}_{p^{\text{\tiny CU}}_{\boldsymbol{\theta}}(\mathbf{c}|\mathbf{s})}\left[\,\sum_{i=1}^{N} \overbrace{\textcolor{black}{ \left\{\mathbb{E}_{p(\mathbf{S},z_i)} \bigg[\,\mathbb{E}_{p^{\text{\tiny SU$_i$}}_{\boldsymbol{\phi}_i}(\mathbf{\hat{x}}_i|\mathbf{c})} [\,\log p(z_i|\mathbf{\hat{x}}_i)]\ \bigg]\,\right\} } }^{\text{Task specific processing}}\right]\ } }_{\text{Cooperative multi-task processing}}.
    \end{aligned}
\end{equation}

To derive the objective function on (\ref{eq:objective_function}), we have approximated the true distribution $p(\mathbf{S},z_i)$ with corresponding available sample set \cite{bishop2006pattern}, and employed the variational method, which is a way to approximate intractable computations based on some adjustable parameters, like weights in NNs \cite{kingma2013auto}. The technique is widely used in machine learning, e.g., \cite{alemi2016deep}, and also in task-oriented communications, e.g., \cite{Shao2021}, \cite{Shao2022}, and \cite{Beck2023}. Thus, our postterior distributions, $p^{\text{\tiny CU}}(\mathbf{c}|\mathbf{s})$ and $p^{\text{\tiny SU}}(\mathbf{x}|\mathbf{c}) = [\,p^{\text{\tiny SU$_1$}}(\mathbf{x}_1|\mathbf{c})\,\dots\,p^{\text{\tiny SU$_N$}}(\mathbf{x}_N|\mathbf{c})]\,$, are approximated by NNs, resulting in $p^{\text{\tiny CU}}_{\boldsymbol{\theta}}(\mathbf{c}|\mathbf{s})$ and $p^{\text{\tiny SU$_i$}}_{\boldsymbol{\phi}_i}(\mathbf{x}_i|\mathbf{c})$, where $\boldsymbol{\theta}$ represents the NN's parameters approximating the CU and $\boldsymbol{\Phi} = [\,\boldsymbol{\phi}_1, \dots, \boldsymbol{\phi}_N ]\,$ is the parameters of the NNs approximating the SU encoders.

As shown in (\ref{eq:objective_function}), by considering the channel outputs we aim to emphasize the role of joint semantic and channel coding performed by our SUs. Employing the fact that $p^{\text{\tiny SU$_i$}}_{\boldsymbol{\phi}_i}(\mathbf{\hat{x}}_i|\mathbf{c})=\int p^{\text{\tiny SU$_i$}}_{\boldsymbol{\phi}_i}(\mathbf{x}_i|\mathbf{c})\,p^{\text{\tiny Channel}}(\mathbf{\hat{x}}_i|\mathbf{x}_i)\,d\mathbf{x}_i$, we try to optimize $p^{\text{\tiny SU$_i$}}_{\boldsymbol{\phi}_i}(\mathbf{\hat{x}}_i|\mathbf{c})$. Moreover, (\ref{eq:objective_function}) shows our adaptation of E2E design, where we jointly optimize the encoders and decoders. Including the AWGN channel directly in our E2E design works well as its transfer function is differentiable. In addition, the outer expectation highlights the difference between our approach and others used for single-task processing and includes cooperation amongst the SU blocks by sharing the CU. This is also explicitly shown by braces in (\ref{eq:objective_function}).

Regarding the i-th decoder in (\ref{eq:objective_function}), the $p^{\text{\tiny Dec$_i$}}(\hat{z}_i|\mathbf{\hat{x}}_i)$ can be fully determined using the known distributions and underlying probabilistic relationship in (\ref{eq:system_probability}) as:
\begin{equation} \label{eq:decoder_probability}
    p^{\text{\tiny Dec$_i$}}(\hat{z}_i|\mathbf{\hat{x}}_i) = \frac{\int p^{\text{\tiny SU$_i$}}_{\boldsymbol{\phi}_i}(\mathbf{\hat{x}}_i|\mathbf{c})\,p^{\text{\tiny CU}}_{\boldsymbol{\theta}}(\mathbf{c}|\mathbf{S})\,p(\mathbf{S},z_i)\,d\mathbf{s}\,d\mathbf{c}}{p(\mathbf{\hat{x}}_i)},
\end{equation}
However, due to the high-dimensional integrals, (\ref{eq:decoder_probability}) becomes intractable and we need to follow the variational approximation technique, resulting in the following:
\begin{equation} \label{eq:objective_function_variational}
    \begin{aligned}
        &\mathcal{L}(\boldsymbol{\theta}, \boldsymbol{\Phi}, \boldsymbol{\Psi}) \approx \\
        &\mathbb{E}_{p^{\text{\tiny CU}}_{\boldsymbol{\theta}}(\mathbf{c}|\mathbf{S})}\left[\,\sum_{i=1}^{N} \left\{\mathbb{E}_{p(\mathbf{S},z_i)} \bigg[\,\mathbb{E}_{p^{\text{\tiny SU$_i$}}_{\boldsymbol{\phi}_i}(\mathbf{\hat{x}}_i|\mathbf{c})} [\,\log q^{\text{\tiny Dec$_i$}}_{\boldsymbol{\psi}_i}(z_i|\mathbf{\hat{x}}_i)]\ \bigg]\,\right\}\right]\,,
    \end{aligned}
\end{equation}
Where in (\ref{eq:objective_function_variational}), $\boldsymbol{\Psi} = [\,\boldsymbol{\psi}_1, \dots, \boldsymbol{\psi}_N]\,$ represents NNs approximating the true distribution of decoders. To obtain the empirical estimate of the above objective function, we approximate the expectations using Monte Carlo sampling assuming the existence of a dataset $\{\mathbf{S}^{(j)}, z^{(j)}_1,\dots,z^{(j)}_N\}^{J}_{j=1}$ for our data-driven approach where $J$ represents the batch size of the dataset.
\begin{equation} \label{eq:empirical_objective}
    \begin{aligned}
        &\mathcal{L}(\boldsymbol{\theta}, \boldsymbol{\Phi}, \boldsymbol{\Psi}) \approx \\[0.6em] &\frac{1}{L}\sum_{l=1}^{L}\left[\,\sum_{i=1}^{N}\left\{\frac{1}{J}\sum_{j=1}^{J}\bigg[\,\frac{1}{K}\sum_{k=1}^{K} [\,\log q^{\text{\tiny Dec$_i$}}_{\boldsymbol{\psi}_i}(\hat{z}_i|\mathbf{\hat{x}}_{j,k})]\,\bigg]\,\right\}\right]\,.
    \end{aligned}
\end{equation}

To overcome the differentiability issues in (\ref{eq:empirical_objective}), we have adopted the reparameterization trick \cite{DBLP:journals/corr/abs-1906-02691}, introducing $\mathbf{c}_{j,l}=\boldsymbol{\mu}_{\mathbf{c}_j} + \boldsymbol{\sigma}_{\mathbf{c}_j} \odot \boldsymbol{\epsilon}_{j,l}$ and $\mathbf{\epsilon}\sim \mathcal{N}(\mathbf{0}, \sigma^2 \mathbf{I})$. The issue of differentiability regarding $\boldsymbol{\phi}$ does not arise, as we presume a deterministic process occurring for $\mathbf{x}_j$, with noise sampling taking place for $\mathbf{\hat{x}}_{j,k} = \mathbf{x}_j + \mathbf{n}_k$. Thus, we fix the sample size of the reparameterization trick, $L$, and the channel sampling size, $K$, to one for each batch. Details on how the objective function is differentiable with respect to all parameters are deferred to Appendix \ref{Appendix:differentiability}.
\vspace{-0.5em}
\section{Simulation Results} \label{section.Simulation}

To demonstrate the effectiveness of our approach, we use the MNIST dataset of handwritten digits \cite{deng2012mnist}, containing 60,000 images for the training set and 10,000 samples for the test set. For a specific number of tasks denoted by $N$, we shape our semantic source as stated before like $\{\mathbf{S}^{(j)}, z^{(j)}_1,\dots,z^{(j)}_N\}^{J}_{j=1}$. This involves pre-processing the MNIST dataset by assigning multiple labels to each data sample $\mathbf{S}^{(j)}$. Thus, in this setup labels stand for our semantic variables. For our evaluations, we consider the execution of two tasks, binary classification (Task1) and categorical classification (Task2). Therefore, two semantic variables, $z_1 \sim Bernoulli$ and $z_2 \sim Multinomial$ will represent Task1 and Task2 respectively. Our experiments consider the classification of digit ``2" as Task1 and digit identification for Task2.
\begin{table}[!t]
    \centering
    \renewcommand{\arraystretch}{1.0}
    \setlength{\tabcolsep}{12pt}
    \caption{The NN structure for MNIST dataset.}
    \begin{tabular}{c|c|c}
        \hline
         & \textbf{Layer} & \textbf{Output size} \\
        \hline
        \textbf{CU} & Fully-connected (FC) + Tanh, FC & $64$ \\
        \hline
        \multirow{2}{*}{\textbf{SU}} & (SU$_1$) FC + Tanh, FC & $16$ \\
        & (SU$_2$) FC + Tanh, FC & $16$ \\
        \hline
        \multirow{2}{*}{\textbf{Dec}} & (Dec$_1$) FC + Tanh, FC + Sigmoid & $1$ \\
        & (Dec$_2$) FC + Tanh, FC + Softmax & $10$ \\
        \hline
    \end{tabular}
    \label{table_implementation}
\end{table}
The implemented NN structure is described in Table \ref{table_implementation} and found heuristically. Regarding the objective function in (\ref{eq:empirical_objective}), the $\sum_{i=1}^{N}$ forces us to feed the dataset in training phase like $\{\mathbf{s}^{(j)}, z^{(j)}_n\}^{J}_{j=1}$. Then, the outer summation will try to capture the learned features across different tasks in the CU, performing a joint training procedure.\\
\begin{figure}[H]
    \centering
    \resizebox{0.43\textwidth}{!}{\input{Figures/bc_error_rates.tikz}}
    \caption{Impact of the CU on task execution for Task1.}
    \label{fig:Task1_cu_WOcu}
\end{figure}
Fig. \ref{fig:Task1_cu_WOcu}, compares the proposed approach (with CU) with the conventional single-task semantic communication (without CU) for Task1, where the task-relevant information extraction and transmission is performed in single stage encoding, just by the SU. It shows that our approach outperforms the conventional one when a statistical relationship exists between tasks, resulting in a lower task execution error rate and faster improvement. The same holds for Task2, as shown in Fig. \ref{fig:Task2_cu_WOcu}. We kept the structure of the SU in ``without CU'' case, the same as our SU and respective decoders in ``with CU'' case for a fair comparison.\\
\begin{figure}[h]
    \centering
    \resizebox{0.43\textwidth}{!}{\input{Figures/cb_error_rates.tikz}}
    \caption{Impact of the CU on task execution for Task2.}
    \label{fig:Task2_cu_WOcu}
\end{figure}

To highlight more benefits of our proposed two-stage semantic encoding, Fig. \ref{fig:Task2_cc_structure} demonstrates that for the single-stage semantic encoding (just SU encoder and decoder) a $4 \times$ larger NN structure is required in comparison to the SU cooperating through the CU.
\begin{figure}[h]
    \centering
    \resizebox{0.43\textwidth}{!}{\input{Figures/cc_structure_error_rates.tikz}}
    \caption{Impact of the CU on simplification of the SUs.}
    \label{fig:Task2_cc_structure}
\end{figure}

Additionally, we illustrate that cooperation through the CU may not always be beneficial, Fig. \ref{fig:constructive_destructive}, contrasts two scenarios: case1 and case2. The proposed approach demonstrates constructive outcomes in case1, involving two categorical classifications. However, in case 2, where both tasks are binary classifications (e.g.,Task1: ``2" and Task2: ``4"), destructive cooperation is evident.\\
\begin{figure}[h]
    \centering
    \resizebox{0.44\textwidth}{!}{\input{Figures/ccbb_error_rates.tikz}}
    \caption{Constructive and destructive behaviors.}
    \label{fig:constructive_destructive}
\end{figure}

We interpret the findings in Fig \ref{fig:constructive_destructive} through the statistical relationship, e.g., Kullback-Leibler (KL) divergence of our semantics distributions, $D_{KL}\left(\,p(z_1|\mathbf{S})\parallel p(z_2|\mathbf{S})\right)\,$. In case1, semantic variables are maximally informative about each other, and constructive cooperation is observed. The converse is observed in case2 where they are less informative. Generally, increased KL divergence corresponds to constructive cooperation, while decreased KL divergence signifies a shift towards destructive behavior.
\vspace{-0.5em}
\section{Conclusion} \label{section.Conclusion}
In conclusion, we investigated a semantic encoding scheme based on an information-theoretic perspective, dividing the encoder into CU and SUs, enabling cooperative semantic communication and the simultaneous process of multiple tasks. We also introduced our semantic source model, allowing the extraction of diverse semantic variables from a single observation. Our findings highlighted that multi-task processing is not always beneficial pointing possible future research directions to us involving optimizing the clustering of SUs semantically to improve cooperative performance and exploring the integration of new SUs into the architecture.

\appendices

\section{Derivation of the Approximated Objective Function} \label{Appendix:derivation}
Using the objective function in (\ref{eq:optimization_problem}):
\begin{equation*}
    \begin{aligned}
        \mathcal{L}(\boldsymbol{\theta}, \boldsymbol{\Phi}) 
        &=\sum_{i=1}^{N}\, I(\mathbf{\hat{x}}_i; z_i) \\[0.6em]
        &= \sum_{i=1}^{N}\, \iint p(\mathbf{\hat{x}}_i, z_i)\,\log \frac{p(z_i|\mathbf{\hat{x}}_i)}{p(z_i)}\,dz_i \, d\mathbf{\hat{x}}_i \\[0.6em]
        &= \sum_{i=1}^{N}\, \left [\ \iint p(\mathbf{\hat{x}}_i, z_i)\,\log p(z_i|\mathbf{\hat{x}}_i)\,dz_i \, d\mathbf{\hat{x}}_i + H(z_i)\, \right ]\ 
    \end{aligned}
\end{equation*}
Further ignoring the constant term ($H(z_i)$) and leverage the Markov chain relationship:
\begin{equation*}
    \begin{aligned}
        \mathcal{L}(\boldsymbol{\theta}, \boldsymbol{\Phi}) &\approx \sum_{i=1}^{N} \iiiiint p(z_i, \mathbf{S})\,p^{\text{\tiny CU}}_{\boldsymbol{\theta}}(\mathbf{c}|\mathbf{S})\\[0.6em]
        &\quad p^{\text{\tiny SU$_i$}}_{\boldsymbol{\phi}_i}(\mathbf{x}_i|\mathbf{c})p^{\text{\tiny Channel}}(\mathbf{\hat{x}}_i|\mathbf{x}_i) \log p(z_i|\mathbf{\hat{x}}_i)\, dz_i\, d\mathbf{s}\, d\mathbf{c}\, d\mathbf{x}_i\, d\mathbf{\hat{x}}_i \\[0.6em]
        &\approx \sum_{i=1}^{N} \iiiint p(z_i, \mathbf{S})\,p^{\text{\tiny CU}}_{\boldsymbol{\theta}}(\mathbf{c}|\mathbf{S})\\[0.6em]
        &\qquad \qquad \qquad p^{\text{\tiny SU$_i$}}_{\boldsymbol{\phi}_i}(\mathbf{\hat{x}}_i|\mathbf{c})\log p(z_i|\mathbf{\hat{x}}_i)\, dz_i\, d\mathbf{s}\, d\mathbf{c}\, d\mathbf{\hat{x}}_i \\[0.6em]
        &\hspace{-1.5em}\approx \mathbb{E}_{p^{\text{\tiny CU}}_{\boldsymbol{\theta}}(\mathbf{c}|\mathbf{s})}\left[\,\sum_{i=1}^{N} \left\{\mathbb{E}_{p(\mathbf{S},z_i)} \bigg[\,\mathbb{E}_{p^{\text{\tiny SU$_i$}}_{\boldsymbol{\phi}_i}(\mathbf{\hat{x}}_i|\mathbf{c})} [\,\log p(z_i|\mathbf{\hat{x}}_i)]\ \bigg]\,\right\}\right]\
    \end{aligned}
\end{equation*}

\section{Differentiability of the Lower Bound} \label{Appendix:differentiability}
Considering the lower bound (\ref{eq:objective_function_variational}):
\begin{equation*} 
    \begin{aligned}
        \mathbb{E}_{p^{\text{\tiny CU}}_{\boldsymbol{\theta}}(\mathbf{c}|\mathbf{s})}\left[\,\sum_{i=1}^{N} \left\{\mathbb{E}_{p(\mathbf{s},z_i)} \bigg[\,\mathbb{E}_{p^{\text{\tiny SU$_i$}}_{\boldsymbol{\phi}_i}(\mathbf{\hat{x}}_i|\mathbf{c})} [\,f(z_i)]\ \bigg]\,\right\}\right]\,
    \end{aligned}
\end{equation*}
We know that $z_i=g(\mathbf{\hat{x}}_i,\boldsymbol{\psi}_i)$, $\mathbf{\hat{x}}_i=h(\mathbf{c},\boldsymbol{\phi}_i,\mathbf{n})$, and $\mathbf{c}=u(\mathbf{s},\boldsymbol{\theta},\boldsymbol{\epsilon})$ where $\mathbf{n}$ in $h(\cdot)$ and $\boldsymbol{\epsilon}$ in $u(\cdot)$ solve the differentiability issues with $\nabla_{\boldsymbol{\theta}} \mathbb{E}_{p^{\text{\tiny CU}}_{\boldsymbol{\theta}}(\mathbf{c}|\mathbf{s})}[\,\cdot]\,$ and $\nabla_{\boldsymbol{\phi}} \mathbb{E}_{p^{\text{\tiny SU}}_{\boldsymbol{\phi}}(\mathbf{\hat{x}}|\mathbf{c})}[\,\cdot]\,$. Employing the chain rule derivative for the lower bound ensures its differentiability with respect to all parameters, as exemplified below for $\boldsymbol{\Phi}$.
\begin{equation*}
    \frac{\mathcal{L}(\boldsymbol{\theta}, \boldsymbol{\Phi}, \boldsymbol{\Psi})}{\boldsymbol{\Phi}} = \frac{\partial f}{\partial g} \cdot \frac{\partial g}{\partial h} \cdot \frac{\partial h}{\partial \boldsymbol{\Phi}}
\end{equation*}

\bibliographystyle{IEEEtran}
\bibliography{IEEEabrv,References.bib}

\end{document}